\documentclass[12pt,preprint]{aastex}

\shorttitle{Point Source Contribution to the Diffuse X-ray Background}
\shortauthors{A.Gupta and M.Galeazzi}

\begin{document}
\title{Contribution of Unresolved Point Sources to the Diffuse X-ray Background below 1 keV}


\author{A. Gupta, and M. Galeazzi\altaffilmark{1}}
\affil{Physics Department, University of Miami, Coral Gables, FL 33124}
\altaffiltext{1}{corresponding author, galeazzi@physics.miami.edu}


\begin{abstract}

We present here the analysis of X-rays point sources detected in several 
observations available in the XMM-Newton public archive. We focused, in 
particular, on energies below 1~keV, which are of particular relevance to 
the understanding of the Diffuse X-ray Background. The average field of all 
the exposures is 0.09 $\textrm{deg}^{-2}$.  We reached an average flux sensitivity of 
$5.8\times10^{-16}~\textrm{erg~s}^{-1}~\textrm{cm}^{-2}$ in the soft 
band (0.5-2.0 keV)  and $2.5\times10^{-16}~\textrm{erg~s}^{-1}~\textrm{cm}^{-2}$ 
in the very soft band (0.4-0.6 keV). In this paper we discuss the logN-logS results, 
the contribution to the integrated X-ray sky flux, and the properties 
of the cumulative spectrum from all sources. In particular, we found 
an excess flux at around 0.5 keV in the composite spectrum of faint 
sources. The excess seems to be a general property of all the fields 
observed suggesting an additional class of weak sources is contributing 
to the X-ray emission at these energies. Combining our results with previous 
investigations we have also quantified the contribution of the 
individual components of the diffuse X-ray Background in the 3/4~keV band.
\end{abstract}



\section{Introduction}

The existence of a Diffuse X-ray Background ($DXB$) was one of the first 
discoveries of extrasolar X-ray astronomy \citep{Giacconi62}. In the 
intervening decades, observations with improving angular and spectral 
resolution have enhanced our understanding of the components that make 
up this background. Above 1~keV, the emission is highly 
isotropic on large angular scales, has extragalactic origin, and can 
be fully accounted by the superposition of unresolved point 
sources \citep{Mushotzky00}. Below 1~keV, the $DXB$ 
is a mixture of the Galactic diffuse emission from a hot bubble 
surrounding the solar neighborhood (the Local Bubble, LB), charge exchange 
from solar wind ions ($SWCX$), hot gas in the galactic halo, and 
an extragalactic flux from point sources and from intercluster 
warm gas that may contain the bulk of the present day 
baryons \citep{Cen99,Ursino06,Galeazzi09}.
Analyses of the soft X-ray background (below 1~keV) usually model 
the Local Bubble component as unabsorbed plasma thermal emission, 
the hotter Galactic halo emission as multiple plasma thermal 
components absorbed by the gas in the Galactic disk, and the unresolved 
extragalactic sources component (primarily active galactic 
nuclei - $AGN$), with an absorbed power law, which is often an 
extension of the power law fit derived at higher energy 
\citep{Gendreau95,McCammon02}.

Recent investigations have attempted to separate the various components 
by using shadow experiments and other observational 
techniques \citep{Galeazzi07, Smith06, Henley08}, however such experiments 
only separate the various components into two major groups, foreground 
and background, and rely on understanding the spectral properties of 
the different components to fully separate them. The contribution of 
unresolved point sources to the $DXB$ dominates above 1~keV and is still 
significant at lower energies. A good understanding of its properties, 
beyond the extension of higher energy investigations, is therefore critical 
for any investigation of the $DXB$. 
This, however, is made difficult by the characteristics of current X-ray 
missions, as high angular resolution missions such as 
Chandra\footnote{http://cxc.harvard.edu/cdo/about$\_$chandra/overview$\_$cxo.html/}, 
designed for good source identification, have a relatively limited 
response in the energy range of interest, while missions designed for 
good response in the low energy range, such as the XQC program 
\citep{McCammon02}, lack the necessary angular resolution. 
XMM-Newton, while not optimized for such an investigation, is a good 
compromise of angular resolution and response below 1~keV, and has been 
used in our investigation.

\section{Data Reduction}

\subsection{Data preparation}

We used the data from 10 observations, corresponding to 5 different targets 
available in the XMM-Newton public archive. The choice of the targets 
was based on several considerations. To limit the effect of absorption 
from the neutral hydrogen and contamination from galactic emission we 
used targets at least $30\degr$ above the galactic plane and with neutral 
hydrogen densities smaller than $2.0\times10^{20}$ cm$^{-2}$. 
To have significant statistics we also limited the investigation to 
targets with at least $80,000$ seconds of good observing time. 
The targets used and their characteristics are summarized in Table~1.

Data from the full XMM-Newton field of view were used in the analysis. 
The raw data were processed using the Standard Analysis Software 
($SAS$)\footnote{http://xmm.vilspa.esa.es/sas/}. Events spread at most in two 
contiguous pixels for the $PN$ (i.e., pattern=0-4) and in four contiguous pixels 
for the $MOS$ (i.e., pattern=0-12) have been selected. Event files were cleaned 
of bad pixels (hot pixels, events out of the field of view, etc.) and 
soft proton flares. The soft proton flares are due to protons with 
energies less than a few hundred keV. The flares can produce a count 
rate up to a factor 100 greater than the mean stationary background 
count rate. They are variable during an observation and from observation 
to observation, and do not have a predictable spectral shape or spatial 
distribution on the detector \citep{Read03}. In order to remove periods 
of unwanted high background levels, we rejected the times with a 0.5-10 keV 
count rate higher than $8~\textrm{counts~s}^{-1}$ for the $PN$ and 
$3~\textrm{counts~s}^{-1}$ for each of the two 
$MOS$ cameras. Multiple observations of the same XMM-Newton targets were 
added using $SAS$ task {\it merge}.

\subsection{Source Detection}

The clean event files were used to generate $MOS1$, $MOS2$, and $PN$ images 
in the 0.4-0.6~keV and 0.5-2.0~keV bands. A corresponding set of exposure 
maps was generated to account for spatial quantum efficiency, mirror 
vignetting, and field of view of each instrument by running XMM-SAS task 
{\it eexpmap}. This task evaluates the above quantities assuming an event 
energy that corresponds to the mean of the energy band boundaries.

The excellent relative astrometry between the three cameras ( within $1''$, 
well under the $FWHM$ of the Point Spread Function - $PSF$) allowed us to merge 
together the $MOS$ and $PN$ images in order to increase the signal-to-noise 
ratio of the sources and reach fainter X-ray fluxes; the corresponding 
exposure maps were merged as well. The source detection and characterization 
procedure applied to the image sets involved the creation of a background 
map for each energy band. As a first step, the XMM-SAS sliding cell 
detection algorithm {\it eboxdetect} was run in local detection mode.  
In this procedure source counts were collected in the cells of 
$5\times5$ pixels adopting a low threshold in the detection likelihood 
to produce a source list. The XMM-SAS task {\it esplinemap} removed from 
the original merged image within a radius of 1.5 times the $FWHM$ of the 
$PSF$ all the sources in the list and creates the so-called $''cheesed''$ 
image and a mask image with value unity in each pixel outside the source 
circles and zero within. We then flat-fielded the cheesed images by 
dividing them by the exposure maps and smoothed them with a Gaussian 
of $5'$ $FWHM$ using $SAS$ task {\it asmooth}. The masked images were also 
smoothed with the same Gaussian as the actual image data. 
Finally, background maps were calculated by dividing the smoothed 
cheese images by the smoothed mask images. The result was multiplied 
by the exposure maps. An example of background map of the Lockman Hole 
is shown in Figure~1. 

Using the calculated background maps, {\it eboxdetect} was run in map 
detection mode to identify point sources. In map detection mode the 
background is taken from the background maps, resulting in improved 
detection sensitivity as compared to the local detection step. 
The source list produced by {\it eboxdetect} was then used as input for 
{\it emldetect}. For all the sources detected with the sliding cell method 
this task performs a maximum likelihood $PSF$ fit. 
In this way refined positions and fluxes for the sources were determined. 
As likelihood threshold for the detection, we adopted the value 
$det_{ml}=6$ (about 3~$\sigma$). Source detection summaries for each pointing 
in both the 0.5-2.0~keV and 0.4-0.6~keV bands are listed in Table~2 
and Table~3 respectively.

The count rate-to-flux conversion factors for individual cameras were 
calculated using $XSPEC$ and the $EPIC$ response matrices generated by 
XMM-SAS tasks {\it rmfgen} and {\it arfgen}. The spectral model used was an 
intrinsic power law $\Gamma=1.52$ affected by Galactic Neutral Hydrogen ($NH$).  
The individual conversion factors are listed in Table~4.  
The total conversion factor ($CF$) was calculated using the exposure 
times for $MOS1$, $MOS2$, and $PN$, and the conversion factors for the three 
instruments, $CFMOS1$, $CFMOS2$, and $CFPN$ following the formula:
\begin{equation}
\frac{T_{TOT}}{CF}=\frac{T_{MOS1}}{CF_{MOS1}}+\frac{T_{MOS2}}{CF_{MOS2}}+\frac{T_{PN}}{CF_{PN}},
\end{equation}
where $T_{TOT}$=($T_{MOS1}$+$T_{MOS2}$+$T_{PN}$). The source flux is then 
straightforward:
\begin{equation}
F_{X}=CF\times CR.
\end{equation}

We generated sensitivity maps for each X-ray field and energy band, 
which contains the faintest flux at which a source can be detected at 
the assumed level of significance above the local background. 
The ``sky coverage'' defines the area of the sky covered down to a given 
flux limit, as a function of the flux. Due to the telescope vignetting 
and the increase in the size of the $PSF$ in the outer regions of the 
detector, the sensitivity decreases toward the outer detector region. 
The sky coverage at a given flux was then obtained by adding up the 
contribution of all detector regions with a given flux limits. 
Figure~2 shows the average sky coverage for all pointing in the 
0.5-2.0~keV and 0.4-0.6~keV bands.

\section{Results}

\subsection{LogN-LogS and Total Flux from Discrete Sources}

The cumulative logN-logS distribution (shown in Figure~3) for all the 
observations has been computed by summing up the contribution of each 
source, weighted by the area in which the source could have been detected, 
following the formula:
\begin{equation}
N (> S) = \sum_{i=1}^{N_s} \left( \frac{1}{\Omega_i}\right) \textrm{deg}^{-2},
\end{equation}
where $N_s$ is the total number of detected sources in the field with 
flux greater than $S$ and $\Omega_{i}$  is the sky coverage associated 
with the flux of the $i^{th}$ source. The variance of the source number 
counts is then defined as:
\begin{equation}
{\sigma_{i}}^{2} = \sum_{i=1}^{N_s} {\left( \frac{1}{\Omega_i}\right)}^{2},           
\end{equation}

Our result in the 0.5-2.0~keV energy band, is qualitatively similar 
to other surveys (e.g., in Figure~3 we show the fits of Giacconi et al. 2001,
Mateos et al. 2008, Hasinger et al. 1998, and Mushotzky et al., 2000), 
however our sample seems consistently richer in faint sources. 
When we fit the data with 
a power law in the flux range $9.0\times10^{-16}$ to 
$1.0\times10^{-13}$ $\textrm{erg~s}^{-1}~\textrm{cm}^{-2}$, the resulting 
best fit to the logN-logS is:
\begin{equation}
N (> S) = (119\pm10) S^{-1.17\pm0.08},
\end{equation}
where $S$ is the flux in units of $10^{-14}~\textrm{erg~s}^{-1}~\textrm{cm}^{-2}$. 

To quantitatively compare our result with previous investigations we looked 
at the number of detected sources at the faint end, bright end, and middle 
point of the flux range covered by our investigation. We compared our 
results both with the previous Chandra surveys by Mushotzky et al. (2000) and 
Giacconi et al. (2001), and the XMM-Newton results by Baldi et al. (2002). 
At the faint end of our logN-logS plot, the number of sources detected by 
our investigation is 42\% higher than the estimates of  Mushotzky (2000) 
and Giacconi (2001) and 35\% higher than the results of Baldi (2002). 
At brighter fluxes our results are consistent, within the errors, with both 
Chandra data and the XMM-Newton survey. At the middle of the logN-logS 
($S=10^{-14}$~cgs), the number of detected sources is 20\% higher than 
the Chandra results and 26\% higher than XMM-Newton results. 
This discrepancy could arise from cosmic variation and/or by the different 
sample investigated. Baldi et al. (2002) used a single, large area of the sky,
while in most other investigations, e.g., Mushotzky (2000) and Hasinger (1998), 
the sources were identified and the logN-logS plot was limited to AGNs. 
While AGNs are expected to be the biggest 
contribution to X-ray point sources, our result, as confirmed by our spectral 
analysis reported in the next section, indicates that there is a significant
component of non-AGN faint sources.

We also computed the resolved intensity of the detected sources by summing 
over the flux of each source divided by the inverse area over which the 
source would have been detected. The total intensity of the sources in 
all the pointings, in the flux range 
$7\times10^{-16}$ to $10^{-14}~\textrm{erg~s}^{-1}~\textrm{cm}^{-2}$ 
for the 0.5-2.0 keV band is 
$4.4\pm0.4\times10^{-12}~\textrm{erg~s}^{-1}~\textrm{cm}^{-2}~\textrm{deg}^{-2}$ 
which corresponds to $\sim$36\% of the total $DXB$ in the same band.

Figure~3 also shows the logN-logS distribution for sources in the 0.4-0.6~keV 
band, extending to a flux limit of 
$5.6\times10^{-12}~\textrm{erg~s}^{-1}~\textrm{cm}^{-2}$. 
Again, we fit the data with a power law in the flux range of 
$5.6\times10^{-16}$ to 
$2.0\times10^{-14}$ $\textrm{erg~s}^{-1}~\textrm{cm}^{-2}$. 
The resulting best fit to the logN-logS is:
\begin{equation}                        
N (> S) = (21.7\pm4.9) S^{-1.16\pm0.15}.
\end{equation}

The total intensity of the sources in all the pointings, in the flux range 
$2.0\times10^{-16}$ to
$2.0\times10^{-14}~\textrm{erg~s}^{-1}~\textrm{cm}^{-2}$ in the 0.4-0.6~keV 
band is 
$1.07\pm0.12\times10^{-12}~\textrm{erg~s}^{-1}~\textrm{cm}^{-2}~\textrm{deg}^{-2}$
which corresponds to $\sim$25\% of the total $DXB$ in that energy band. 
We note that all the sources identified in the 0.4-0.6~keV band are also 
identified in 0.5-2.0~keV band.

\subsection{Energy spectra}

We measured the average stacked spectrum of all the sources detected in the
0.5-2.0 keV and 0.4-0.6 keV bands. For this purpose, we use only $PN$ data. 
The background spectrum was obtained using the event file of the total field 
of the same exposure, after removing the detected sources. The background 
was scaled by the ratio of the total exposure maps of the sources and the 
background. The $SAS$ tasks {\it rmfgen} and {\it arfgen} were used to 
produce $EPIC$ response matrices.
We used $XSPEC$\footnote{http://heasarc.gsfc.nasa.gov/docs/xanadu/xspec/manual/manual.html}  
to compute the slope of a power law spectrum with average 
$NH$=$1\times10^{-20}~\textrm{cm}^{-2}$. The fit was over the energy 
range 0.5-7.5 keV (the $PN$ spectrum shows a strong Cu-K line at 8.1~keV 
that should be avoided). We obtained a photon index of $1.75\pm0.07$ 
with normalization $7.24\pm0.07$ 
$\textrm{photons~keV}^{-1}~\textrm{s}^{-1}~\textrm{cm}^{-2}~\textrm{sr}^{-1}$ 
and $1.93\pm0.08$ with normalization of $5.87\pm0.06$ 
$\textrm{photons~keV}^{-1}~\textrm{s}^{-1}~\textrm{cm}^{-2}~\textrm{sr}^{-1}$  
for  all the sources detected in the 0.5-2.0 keV and 0.4-0.6 keV band 
respectively. Errors refer to 90\% confidence levels. Both spectra with 
the best fit power law are shown in Figure~4. We note that, as it may be 
expected, the spectrum of the sources identified in the 0.4-0.6 keV 
energy band is softer than that for the sources identified in the 0.5-2 keV 
band. The spectrum of the sources identified in the 0.5-2.0 keV band shows
an excess of counts around 1 keV at the three sigma level ($\chi ^2=26.7$, $n=9$). 
The excess is well fitted with a zero redshift thermal component with 
temperature $T=0.92$~keV, and emission measure $EM=9\times 10^{-5}$~cm$^{-6}$~pc, 
corresponding to a 
flux of $6.5\times 10^{-14}$~ergs~s$^{-1}$~cm$^{-2}$~deg$^{-2}$. 
We attribute this thermal component 
to the contribution of stars in the Milky Way \citep{Kashyap92}.

We also investigated the dependence of the spectral shape on the source 
brightness. This is particularly important as in most studies of the $DXB$ only 
bright sources are detected and their properties are extended to faint 
ones. For this purpose we divided the detected sources into two groups, 
bright and faint, based on their flux. The threshold between the two 
groups was set to $2.0\times10^{-15}~\textrm{erg~s}^{-1}~\textrm{cm}^{-2}$. 
The spectrum of bright sources detected in the 0.5-2.0 keV band was well fitted 
in the energy range 0.5-7.5 with a power law of photon index $1.87\pm0.07$ and 
normalization $6.39\pm0.06$ 
$\textrm{photons~keV}^{-1}~\textrm{s}^{-1}~\textrm{cm}^{-2}~\textrm{sr}^{-1}$ 
plus the $T=0.92$~keV thermal component discussed above (see Figure~5). 
However, the spectrum of faint sources has an excess of counts below 
0.7~keV and could not be fitted with a simple power law in that range (Figure~5). 
We performed a power law fit over the energy range 0.7-7.5 keV, obtaining 
a photon index $\Gamma=1.05\pm0.03$ with normalization $0.70\pm0.02$ 
$\textrm{photons~keV}^{-1}~\textrm{s}^{-1}~\textrm{cm}^{-2}~\textrm{sr}^{-1}$.
Notice that the spectrum of faint sources does not show any significant 
thermal component around 1~keV.

We investigated the nature of the excess counts below 0.7 keV and here is a 
summary of our conclusions:
\begin{itemize}
\item The excess has a statistical significance of several sigmas 
($\chi^{2}=217$, $n=39$) and cannot be simply explained by statistical
fluctuations.
\item The excess was present in all pointings and is not due to a single 
anomalous source or target.
\item We tried to fit the spectrum of faint sources with two power laws, 
one dominating above 1~keV, the other below 1~keV, but we could not improve 
the fit significantly.
\item We also tried to fit the spectrum with a power law plus a thermal 
component (Figure~6) and the fit was greatly improved. Notice that, since 
the spectrum is the sum of the contribution from several sources, a fit 
with a single thermal component has, per se, limited significance. 
However, the goodness of the fit is a strong indication that the excess flux 
may be thermal in nature.
The best fit parameter for the thermal component are $T=2.1\times10^{6}$~K, 
redshift $z=0.02$, and emission measure $EM=0.00014$~cm$^{-6}$~pc, for a total
flux of the excess component equal to 
$8.5\times 10^{-14}$~ergs~s$^{-1}$~cm$^{-2}$~deg$^{-2}$.
The corresponding power law has photon index $\alpha=(1.03\pm0.02)$ and 
normalization 
$0.55\pm0.02$ $\textrm{photons~keV}^{-1}~\textrm{s}^{-1}~\textrm{cm}^{-2}~\textrm{sr}^{-1}$.
At this temperature the thermal component is expected to have a significant emission in
O~{\tiny VII} and O~{\tiny VIII}, at a redshift that is practically indistinguishable
from zero with the resolution of current satellites. The estimated
emission from the excess thermal component is  $0.19$~LU 
(photons~s$^{-1}$~cm$^{-2}$~sr$^{-1}$) 
and $0.06$~LU for the O~{\tiny VII} and O~{\tiny VIII} respectively.

\item We investigated the possibility that the excess is due to the contribution 
of Milky Way stellar sources. Kashyap et al. (1992) estimated the contribution 
to the diffuse soft X-ray background flux from Galactic stars at various 
energies ranging from 0.1 to $\sim$5~keV. They found that stellar contribution 
at high Galactic latitudes is less than 3\% for photon energies less than 
0.3~keV, 3\%-17\% in the medium energy ($\sim$0.4 to $\sim$0.9~keV) and 
10\%-30\% in the high energy band ($\sim$0.8-2.0~keV). 
Stars mainly have two temperature thermal spectra, since stellar corona emission is 
composed of two components namely a hot active component at nominal 
temperature $kT_a\sim 1$~keV ($1.1 \times 10^{7}$~K) and a somewhat cooler 
quiescent component $kT_q\sim 0.3$~keV (($3.5 \times 10^{6}$~K). 
In some stars only one thermal component is present, with a mean coronal 
temperature $kT$ in the range [0.5, 0.8~keV] ($5.8-9.3 \times 10^{6}$~K; 
see e.g. Kashyap et al. 1992, Della Ceca et al. 2004, Lopez-Santiago et al. 2007).
We have already discussed the excess thermal emission at $T=0.92$~keV which we
attribute to the stellar hot active component and we investigated the 
possibility that this lower temperature excess could be due to the cooler quiescent 
component.
We fitted the excess with a thermal component at redshift zero and 
obtained a good fit with temperature $kT=0.15$~keV ($1.7 \times 10^{6}$~K) 
and emission measure $EM=0.00016$~cm$^{-6}$~pc.
The total flux of the excess thermal component is compatible with what expected for stellar
contribution, however, the best fit temperature is significantly smaller than
the temperature range predicted by Kashyap et al. (1992).
We also tried to fit the excess emission constraining the temperature values 
to the ranges described by Kashyap et al. (1992), using both one and 
two-temperature models, but we were unable to obtain a good fit.
\end{itemize}

In conclusion, we believe that, when we look at faint sources, in addition 
to the typical $AGN$ contribution, there is also a significant contribution 
from a different class of sources with primarily thermal emission. 
Although the total flux of this contribution is consistent with stellar origin,
its temperature is too low compared to typical stellar emission.
This class of sources may also be explained by unresolved faint galaxies, galaxy 
clusters, and/or groups, and must be taken into account in the $DXB$ budget.

We repeated the same investigation for sources detected in the energy 
band 0.4-0.6 keV.  The spectrum of bright 
(flux $\geq$ $10^{-15}~\textrm{erg~s}^{-1}~\textrm{cm}^{-2}$) 
sources detected in the 0.4-0.6 keV band is well described with a 
power law fit of photon indexes $2.05\pm0.02$ with normalization of 
$4.74\pm0.04$ 
$\textrm{photons~keV}^{-1}~\textrm{s}^{-1}~\textrm{cm}^{-2}~\textrm{sr}^{-1}$
(Figure 7). The spectrum of faint sources has a $3\sigma$ statistical 
significance for the thermal excess below 0.7 keV and could not be fitted 
with a simple power law in that range (Figure~7). We performed the power 
law fit over the energy range 0.7-7.5 keV, obtaining a photon 
index $\Gamma=1.69\pm0.02$ with normalization of $1.02\pm0.03$ 
$\textrm{photons~keV}^{-1}~\textrm{s}^{-1}~\textrm{cm}^{-2}~\textrm{sr}^{-1}$. 
In Figure~8 we show the spectrum of faint sources with overlapped the power 
law plus thermal component model. For the thermal component we used the same 
parameters as for the fit of faint sources identified in the soft band. 
The result of the spectral analysis in both bands is summarized in 
Tables~5 and 6.

\section{The Diffuse X-ray Background}

We combined our result with previous results to create
a global picture of the $DXB$ emission below 1~keV. 
As discussed before, in this energy range, the $DXB$ is a mixture of the Galactic 
local bubble ($LB$) emission, solar wind charge exchange ($SWCX$), 
hot gas in the galactic halo ($GH$), and 
extragalactic flux from point sources and from intergalactic 
warm-hot gas ($WHIM$).

The ROSAT All Sky Survey ($RASS$) represents the most extensive study of the
X-ray diffuse emission below 1~keV. 
Analysis of $RASS$ data and subsequent observations has shown that there is significant 
spatial and temporal variation in 
the $DXB$ emission \citep{Snowden00}. The temporal variation is mostly attributable to the $SWCX$, 
while the spatial variation seems mostly due to $LB$ and $GH$ emission. 
An average over the whole sky indicates that the total diffuse X-ray 
emission in the $3/4$~keV energy band is 
$3.04\times 10^{-12}$~ergs~s$^{-1}$~cm$^{-2}$~deg$^{-2}$ \citep{kuntz01}.
More recent work using XMM-Newton, Chandra, and Suzaku produces similar results.
In particular, combining observations performed by Smith et al. (2005, 2006), Galeazzi et al. (2007), 
Henley et al. (2007, 2008), and Gupta et al. (2009) we obtained a total flux in 
O~{\tiny VII} plus O~{\tiny VIII} lines of 
$(3.1\pm 0.8)\times 10^{-12}$~ergs~s$^{-1}$~cm$^{-2}$~deg$^{-2}$
($12.5\pm 3.6$~LU). 
Considering that the $3/4$~keV band is dominated by these lines,
the two results are in good agreement. 

Moreover, the observations by Smith et al. (2005, 2006), Galeazzi et al. (2007), 
Henley et al. (2007, 2008), and Gupta et al. (2009) 
are shadow experiments and can be used to separate the $DXB$ emission
into two components, a foreground component consisting of $LB$ and
$SWCX$ emission, and a background component consisting of $GH$,
point sources, and $WHIM$.
A typical shadow experiment consists of two observations, one 
in the direction of a high neutral hydrogen density cloud at a distance of 
$50-200$~pc, the other in the direction of a low neutral hydrogen column 
density region as close as possible to the cloud. As the cloud absorbs most of 
the background X-ray emission, the comparison of the two observations 
allows a clean separation between foreground  and background.
For example, in our previous investigation \citep{Galeazzi07} we used the high 
density, high latitude, neutral hydrogen cloud MBM20 which absorbs about 
75\% of the background in the energy range of 0.4-0.6~keV and a low-density region 
nearby, the Eridanus Hole, which absorbs only 8\% of the background 
in the same energy range.
Similarly, Smith et al. (2005, 2006) performed shadow observations in the direction of 
the cloud MBM12, and Henley et al. (2007, 2008) in the direction of a filament 
in the southern galactic emisphere.
Combining all shadow investigations to account for spatial and temporal variations
in the total diffuse flux, we found that ($30\pm 12$)\% (or $3.75\pm 1.3$~LU in 
oxygen lines) of the diffuse emission is due to foreground sources, while ($70\pm 12$)\% 
(or $8.75\pm 1.3$~LU in oxygen lines) is due to background sources. 

Using the point sources identified in this investigation we can also set a lower limit
to the diffuse emission due to usually unidentified point sources. 
Scaled to the $3/4$~keV band, the cumilative flux of all point sources
identified in this investigation corresponds to ($35\pm 11$)\% of the 
total $DXB$ emission. 
We recently also performed an investigation of the $WHIM$ emission using 
a statistical approach on the same targets used in this investigation \citep{Galeazzi09}.
In our investigation we found clear evidence of the emission from the 
$WHIM$ and we quantified it as ($12\pm 5$)\% of the total $DXB$ in the 
energy band 0.4-0.6~keV. A summary of the results with the contribution 
from each component is shown in Table~7. 

\section{Conclusions}

We investigated the properties of point sources using data from the 
XMM-Newton public archive, focusing, in particular, on the properties 
below 1 keV and the contribution to the Diffuse X-ray Background. We 
looked at sources identified in two separate energy bands, a typical 
soft band 0.5-2 keV and a very soft one 0.4-0.6 keV. In the soft band, 
the sources detected in all the pointings at fluxes from 
$7.0\times10^{-16}~\textrm{erg~s}^{-1}~\textrm{cm}^{-2}$ to 
$1.0\times10^{-13}~\textrm{erg~s}^{-1}~\textrm{cm}^{-2}$  contribute a 
total flux of 
$(4.4\pm0.4)\times10^{-12}~\textrm{erg~s}^{-1}~\textrm{cm}^{-2}~\textrm{deg}^{-2}$,
which corresponds to $\sim$36\% of the total CXB. The flux resolved in the very 
soft band from  
$2\times10^{-16}$ to $2\times10^{-14}~\textrm{erg~s}^{-1}~\textrm{cm}^{-2}$ 
contributes a flux of 
$(1.07\pm0.12)\times10^{-12}~\textrm{erg~s}^{-1}~\textrm{cm}^{-2}~\textrm{deg}^{-2}$  
corresponding to $\sim$25\% of the total CXB. 

We obtained the cumulative spectra of all the sources identified in the two 
energy bands and also classified our sources using different flux thresholds. 
The threshold between the two groups was set to $2.0\times10^{-15}$ and 
$10^{-15}~\textrm{erg~s}^{-1}~\textrm{cm}^{-2}$ in the soft band and very 
soft band respectively.
A power law fit of the spectrum from bright sources in the range
0.5-7.5~keV using the average Galactic 
value of $NH=1\times10^{-20}~\textrm{cm}^{-2}$ yields photon indexes of  
$1.77\pm0.01$ and $2.05\pm0.02$ for the sources detected in the soft band and very 
soft band respectively. When looking at faint sources, we found an excess flux 
around 0.5 keV in both bands that seems to be thermal in nature. 
We attribute the excess to a class of sources different from the typical 
$AGN$ component, possibly unresolved galaxy clusters and/or groups, or 
coronal emission from stars in the Milky Way.

\acknowledgments

We would like to thank Dan McCammon and Steve Snowden for the very useful 
discussion and suggestions.

\clearpage

\begin{figure}
\includegraphics[height=8cm]{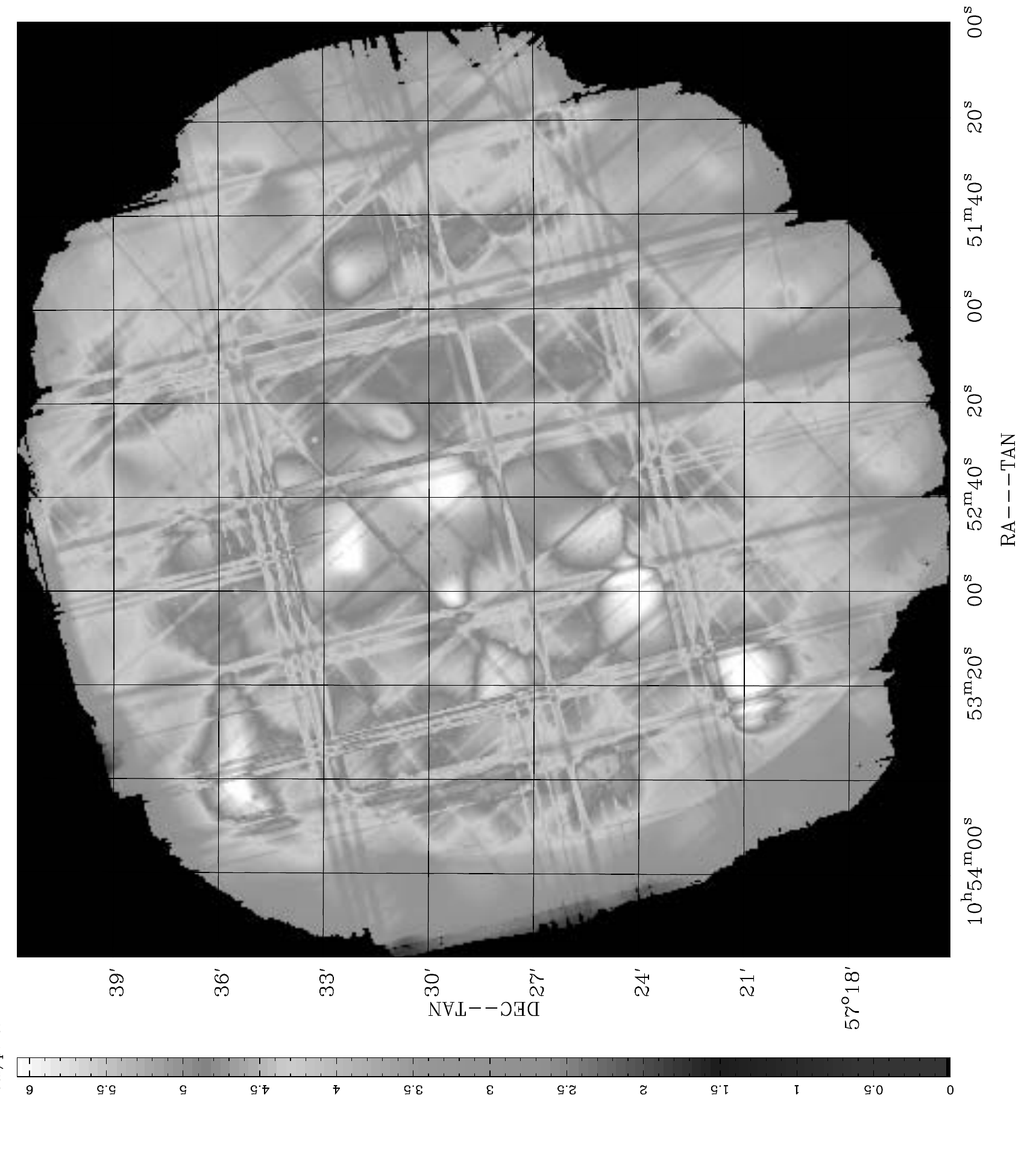}
\caption{Background map derived as described in the text for the Lockman Hole target.}
\label{fig1}
\end{figure}

\clearpage

\clearpage

\begin{figure}
\includegraphics[height=8cm]{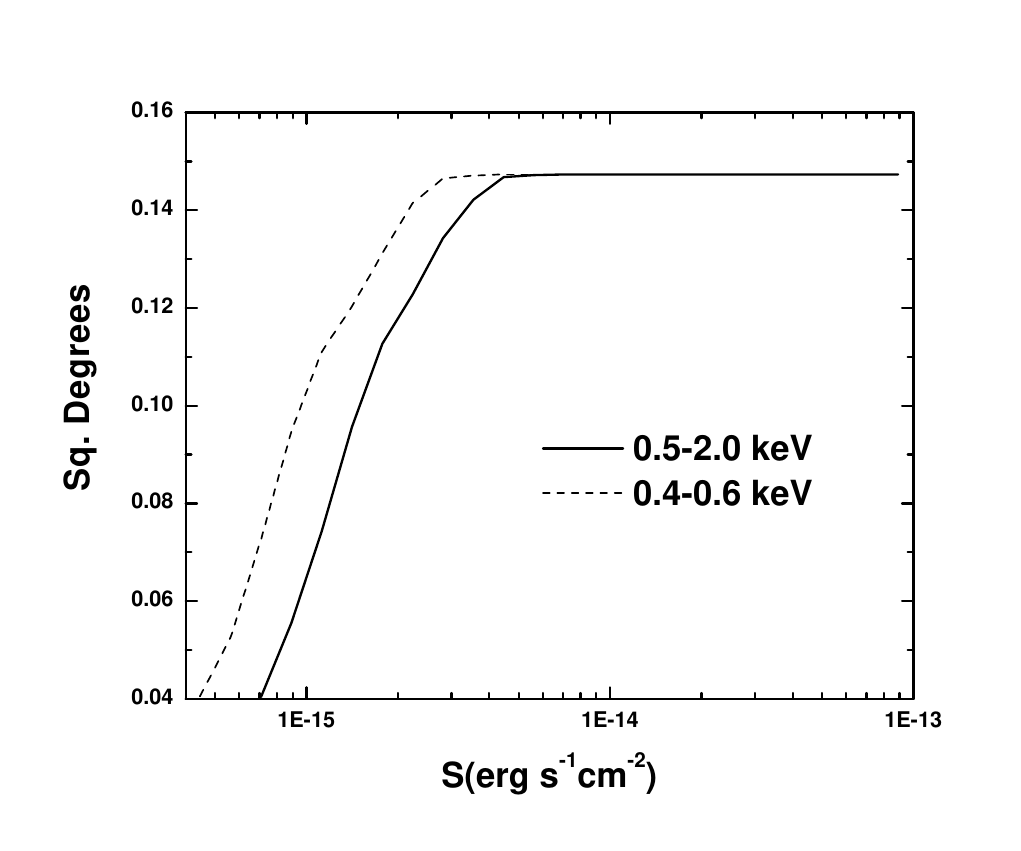}
\caption{Average sky coverage for all pointing in the 0.5-2.0 keV and 0.4-0.6 keV bands.}
\label{fig2}
\end{figure}

\clearpage

\clearpage

\begin{figure}
\includegraphics[height=8cm]{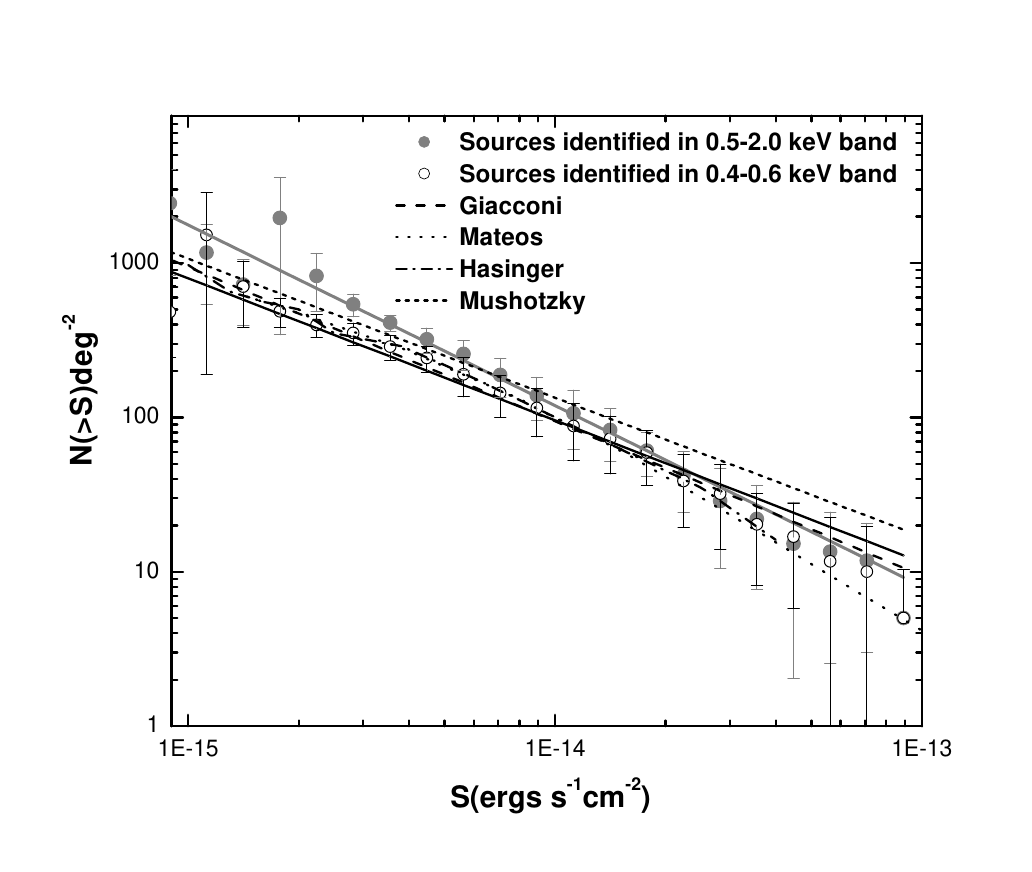}
\caption{The average logN-logS in the 0.5-2 keV (full circles) and 0.4-0.6 keV 
(empty circles) bands for all targets used in this investigation. The black and gray 
curves represent the power law fit to the two experimental data sets respectively. 
The dotted line represents the best fit from Giacconi et al. (2001).}
\label{fig3}
\end{figure}

\clearpage

\clearpage

\begin{figure}
\includegraphics[height=8cm]{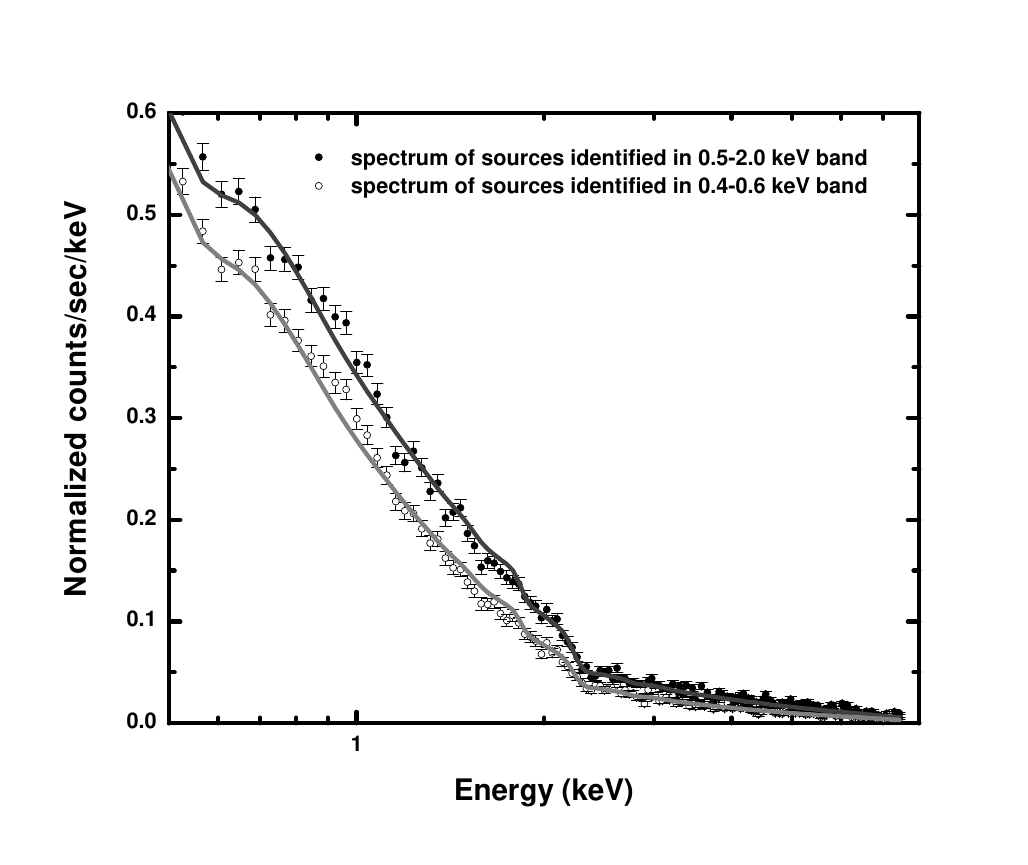}
\caption{Average spectra of all the sources detected in the 0.5-2.0 keV (full black 
circles) and 0.4-0.6 keV (empty grey circles) bands. The dark grey and grey curve represents 
the power law fit in energy range 0.5-7.5 keV, for the two bands respectively.}
\label{fig4}
\end{figure}

\clearpage

\clearpage

\begin{figure}
\includegraphics[height=8cm]{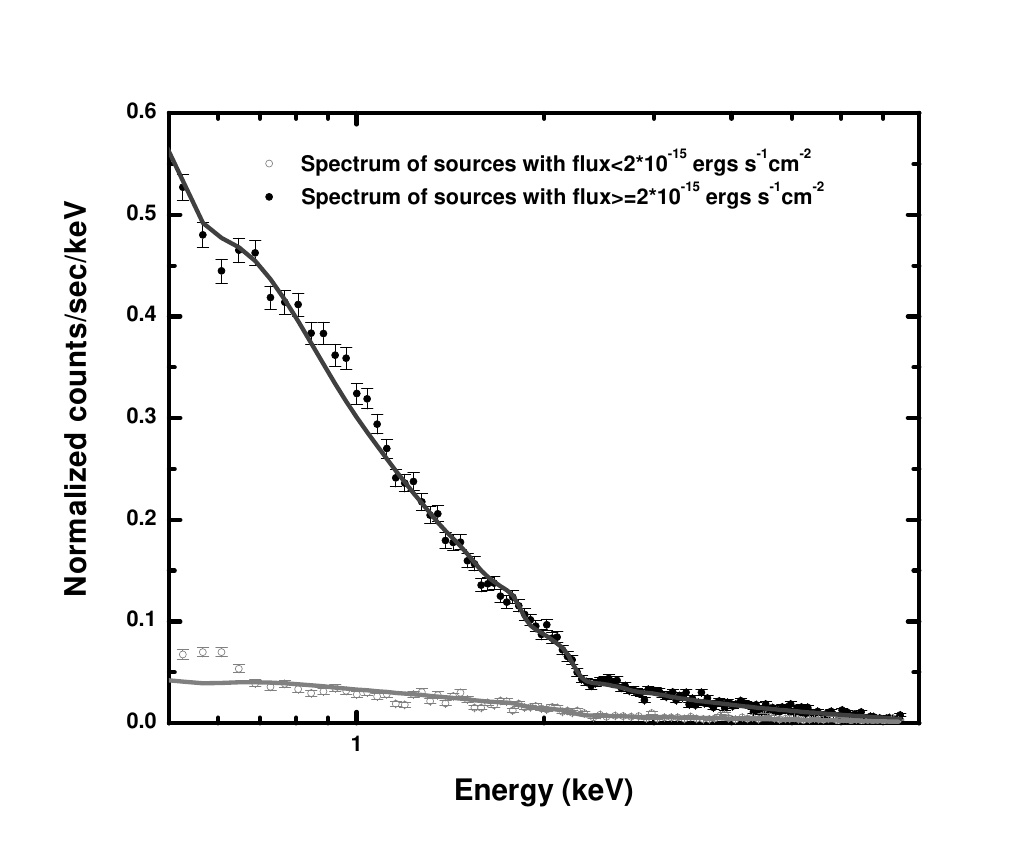}
\caption{The average spectra of bright (full black circles) and faint (empty grey circles) 
sources detected in the 0.5-2.0 keV band. The dark grey and grey curves represent the power 
law fits for the bright and faint sources respectively. Note the soft excess at 0.5-0.65 keV for the faint sources.}
\label{fig5}
\end{figure}

\clearpage

\clearpage

\begin{figure}
\includegraphics[height=8cm]{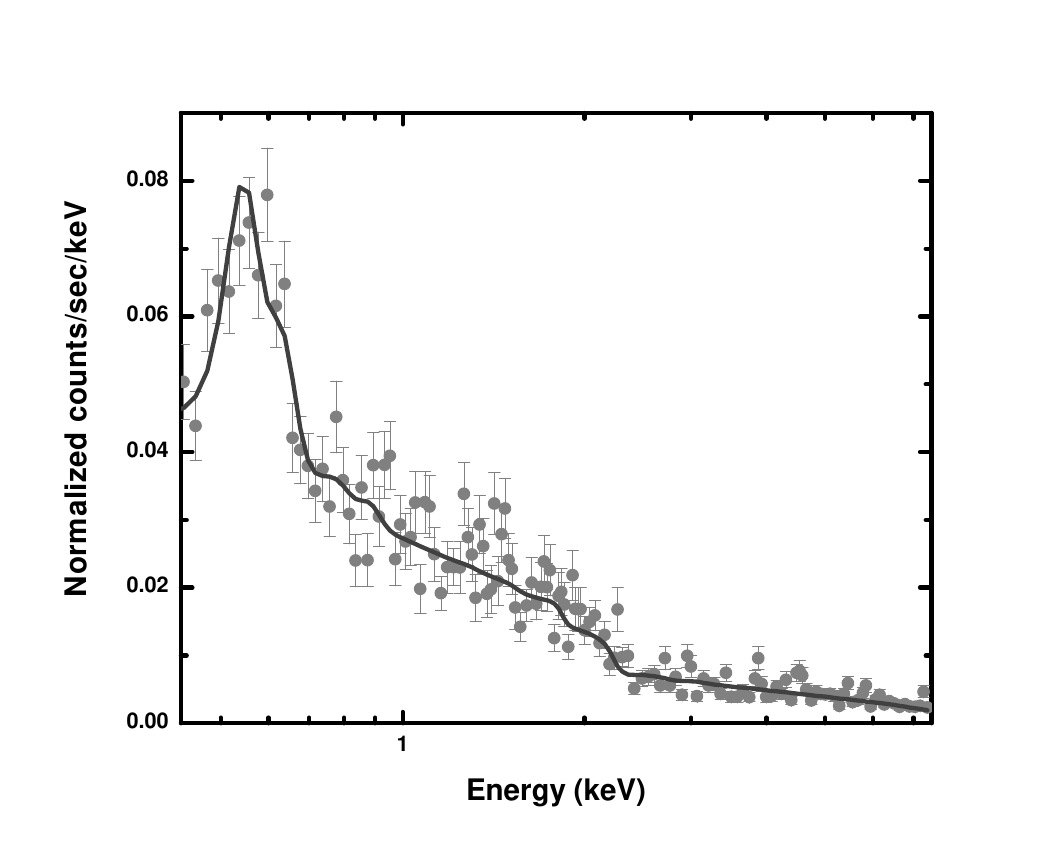}
\caption{Average spectrum of faint sources fitted with an absorbed power law plus 
thermal component.}
\label{fig6}
\end{figure}

\clearpage

\clearpage

\begin{figure}
\includegraphics[height=8cm]{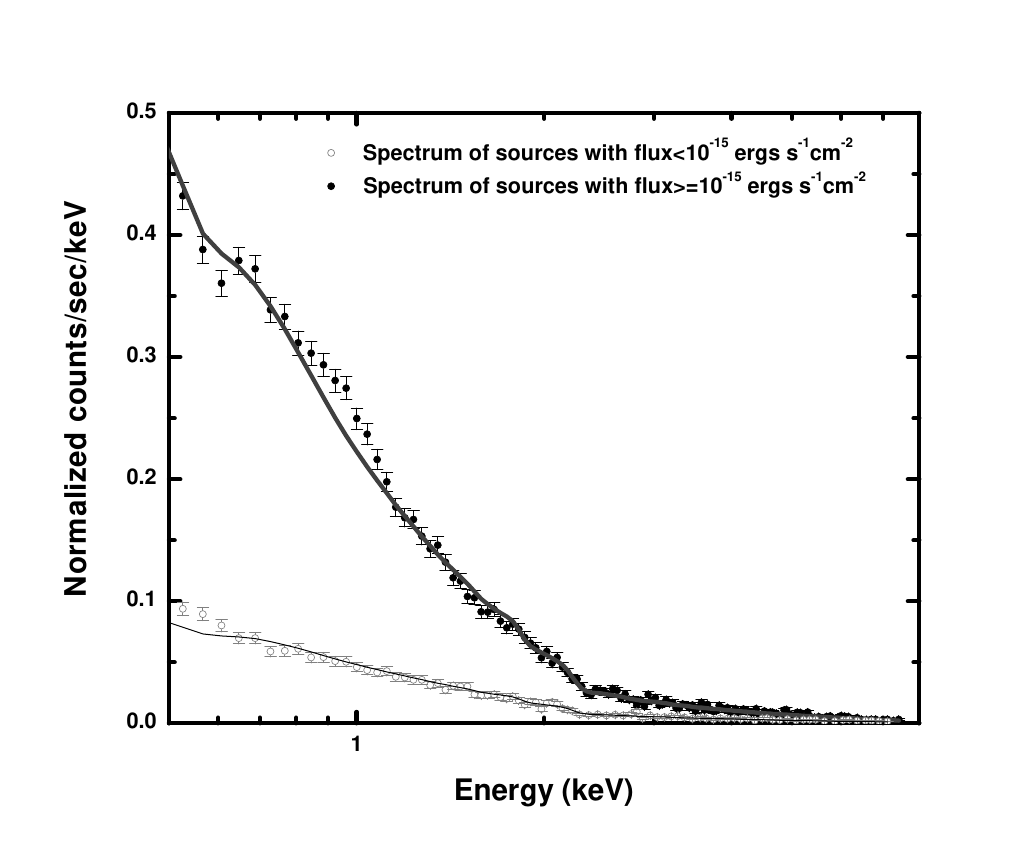}
\caption{The average spectra of bright (full black circles) and faint (empty grey circles) 
sources detected in the 0.4-0.6~keV band. The dark grey and grey curves represent the power 
law fits for the bright and faint sources respectively.}
\label{fig7}
\end{figure}

\clearpage

\clearpage

\begin{figure}
\includegraphics[height=8cm]{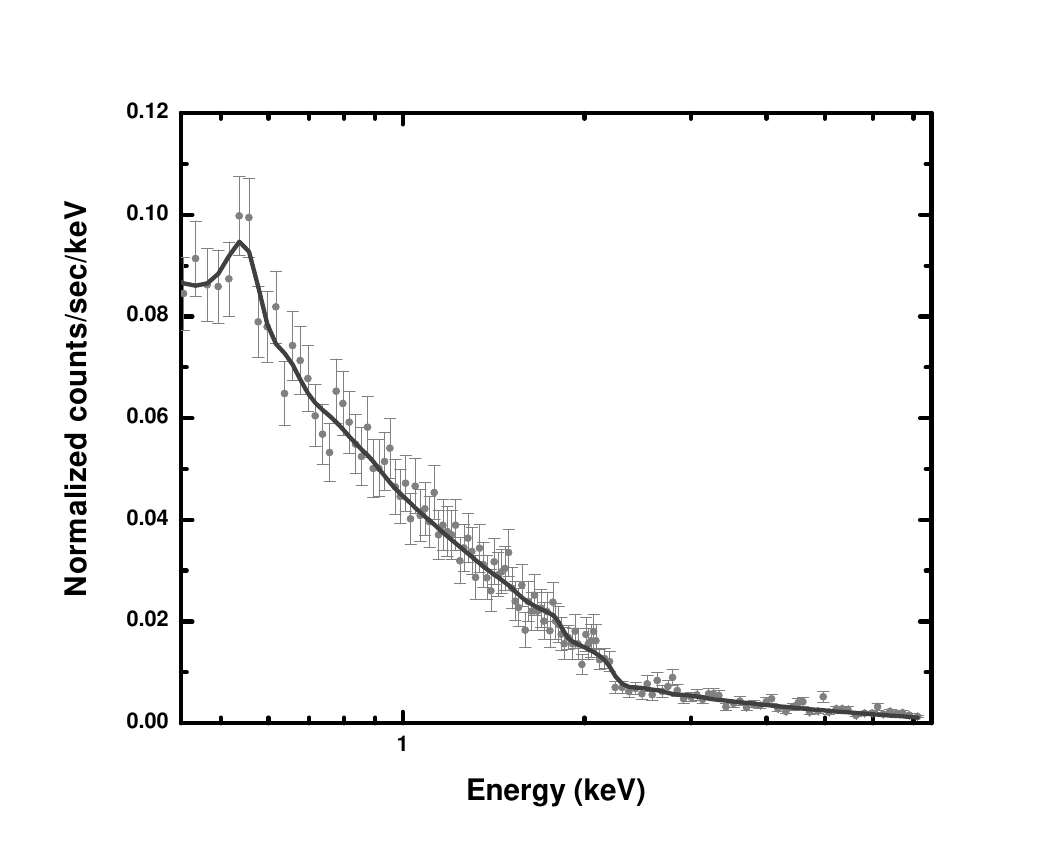}
\caption{Average spectrum of faint sources fitted with an absorbed power law plus 
thermal component.}
\label{fig8}
\end{figure}

\clearpage

\begin{deluxetable}{lccc}
\tablewidth{0pc}
\tablecaption{Targets used in this investigation.}
\tablehead{
\colhead{Target}           & \colhead{l}      &
\colhead{b}          & \colhead{NH($10^{20}$ $cm^{-2}$)}}
\startdata
Lockman Hole & 149 16 48.1 & 53 08 45.9 & 0.7\\
Hubble Deep Field N & 125 53 31.3 & 54 48 52.4 & 1.4\\
Deep Field 1334+37 & 85 37 22.5 & 75 55 16.4 & 0.8\\
Eridanus Hole & 213 25 17.9 & -39 04 25.6 & 0.86\\
AXAF Ultra deep F & 223 34 36 & -54 26 33.3 & 1.0\\
\enddata
\end{deluxetable}

\clearpage

\begin{deluxetable}{lcccccc}
\tablecolumns{6}
\tablewidth{0pc}
\tablecaption{Source detection in the 0.5-2.0 keV band.}
\tablehead{
\colhead{Target}    & \colhead{Exposure Time\tablenotemark{a}}  & \colhead{Minimum  flux}& \multicolumn{3}{c}{Sample}\\
\cline{4-6}\\
\colhead{} & \colhead{sec} & \colhead{$\textrm{erg~s}^{-1}\textrm{cm}^{-2}$} & \colhead{Total} & 
\colhead{Faint\tablenotemark{b}} & \colhead{Bright\tablenotemark{c}}
}
\startdata
\textbf{Lockman Hole} & 628945 & $2.34\times10^{-16}$ & 291 & 185 & 106\\
\textbf{Hubble Deep Field} & 131751 & $7.18\times10^{-16}$ & 144 & 57 & 87\\
\textbf{Deep Field 1334+37} & 165989 & $6.45\times10^{-16}$ & 159 & 60 & 99\\
\textbf{Eridanus Hole} & 50230 & $9.96\times10^{-16}$ & 110 & 38 & 72\\
\textbf{AXAF Ultra deep F} & 431618 & $31.9\times10^{-16}$ & 225 & 149 &76\\
\enddata

\tablenotetext{a}{Total exposure time from all available observations of the same target.}
\tablenotetext{b}{sources with flux$~\leq$ $2.0\times10^{-15}~\textrm{erg~s}^{-1}~\textrm{cm}^{-2}$}
\tablenotetext{c}{sources with flux$~\geq$ $2.0\times10^{-15}~\textrm{erg~s}^{-1}~\textrm{cm}^{-2}$}

\end{deluxetable}

\clearpage

\begin{deluxetable}{lccccc}
\tablecolumns{5}
\tablewidth{0pc}
\tablecaption{Source detection in the 0.4-0.6 keV band.}
\tablehead{
\colhead{Target}     & \colhead{Minimum detected flux} & \multicolumn{3}{c}{Sample}\\
\cline{3-5}\\
\colhead{}  & \colhead{$\textrm{erg~s}^{-1}\textrm{cm}^{-2}$} & \colhead{Total} & \colhead{Faint\tablenotemark{a}} & \colhead{Bright\tablenotemark{b}}}
\startdata
\textbf{Lockman Hole}  & $1.67\times10^{-17}$ & 143 & 99 & 44\\
\textbf{Hubble Deep Field} & $4.02\times10^{-16}$ & 50 & 13 & 37\\
\textbf{Deep Field 1334+37} & $3.84\times10^{-16}$ & 75 & 28 & 47\\
\textbf{Eridanus Hole} & $4.48\times10^{-16}$ & 47 & 15 & 32\\
\textbf{AXAF Ultra deep F}  & $1.6\times10^{-17}$ & 108 & 78 & 30\\
\enddata

\tablenotetext{a}{sources with flux $\leq$ $10^{-15}~\textrm{erg~s}^{-1}~\textrm{cm}^{-2}$}
\tablenotetext{b}{sources with flux $\geq$ $10^{-15}~\textrm{erg~s}^{-1}~\textrm{cm}^{-2}$}

\end{deluxetable}

\clearpage

\begin{deluxetable}{lcc}
\tablewidth{0pt}
\tablecaption{Count-rate-to-flux conversion factors} 
\tablehead{
\colhead{EPIC Camera}   & \colhead{0.5-2.0 keV}   &  \colhead{0.4-0.6 keV}}
\startdata
$PN$ & $1.85\times10^{-12}$ & $1.48\times10^{-12}$\\
$MOS$ & $5.9\times10^{-12}$ & $6.18\times10^{-12}$\\
\enddata

\tablecomments{Count-rate-to-flux conversion factors for the individual EPIC cameras 
and energy bands, stated in units of $\textrm{erg~s}^{-1}~\textrm{cm}^{-2}$ for a 
rate of 1  $\textrm{counts~s}^{-1}$. A photon-index power law of $\Gamma =1.52$ affected 
by Galactic absorption of $1.0\times10^{20}~\textrm{cm}^{-2}$ was assumed. 
Both $MOS$ cameras were assumed to be identical.}

\end{deluxetable}

\clearpage

\begin{deluxetable}{lccccc}
\tablewidth{0pt}
\tablecaption{Spectral Fits on the 0.5-2.0 keV energy range}
\tablehead{
\colhead{Sample} & \colhead{Objects}   &  \colhead{$\Gamma$\tablenotemark{a}} & \colhead{Flux\tablenotemark{b}} & \colhead{Norm\tablenotemark{c}} & \colhead{Reduced $\chi^{2}$}}
\startdata
Total & 929 & $1.77\pm0.01$ & $4.8\times10^{-12}$ & $7.24\pm0.07$ & 1.3\\
Faint & 489 & $1.05\pm0.03$ & $3.8\times10^{-13}$ & $0.70\pm0.02$  & 1.9\\
Bright & 440 & $1.87\pm0.01$ & $4.8\times10^{-12}$ & $6.39\pm0.06$  & 1.1\\
\enddata

\tablenotetext{a}{Power law index of photon spectrum fit in the  energy range of 0.5-7.5 keV}
\tablenotetext{b}{Flux of the sources in the energy range of 0.5-2.0 keV in units of  $\textrm{erg~s}^{-1}\textrm{cm}^{-2}\textrm{deg}^{-2}$}
\tablenotetext{c}{Normalization of power law fit in units of $\textrm{photons~keV}^{-1}~\textrm{s}^{-1}~\textrm{cm}^{-2}~\textrm{sr}^{-1}$}

\end{deluxetable}

\clearpage

\begin{deluxetable}{lccccc}
\tablewidth{0pt}
\tablecaption{Spectral Fits on the 0.4-0.6 keV energy range}
\tablehead{
\colhead{Sample} & \colhead{Objects}   &  \colhead{$\Gamma$} & \colhead{Flux\tablenotemark{a}} &\colhead{Norm} & \colhead{Reduced $\chi^{2}$}}
\startdata
Total & 423 & $1.93\pm0.01$ & $1.1\times10^{-12}$ & $5.87\pm0.06$ &  1.1\\
Faint & 233 & $1.69\pm0.04$ & $1.5\times10^{-13}$ & $1.02\pm0.03$ & 0.89\\
Bright & 190 & $2.05\pm0.02$ & $8.6\times10^{-13}$ & $4.74\pm0.04$ & 0.94\\
\enddata

\tablenotetext{a}{Flux of the sources in the energy range of 0.4-0.6 keV in units of  $\textrm{erg~s}^{-1}\textrm{cm}^{-2}\textrm{deg}^{-2}$}

\end{deluxetable}

\clearpage

\begin{deluxetable}{lc|lc}
\tablewidth{0pt}
\tablecaption{Components of the total diffuse X-ray emission in the $3/4$~keV energy band}
\tablehead{\multicolumn{4}{c}{Total Luminosity = 6.25~keV~s$^{-1}$~cm$^{-2}$~sr$^{-1}$}}
\startdata
Foreground & ($30\pm 12$)\% & SWCX \\
           &                & Local Bubble \\
\hline
Background & ($70\pm 12$)\% & Galactic Halo \\
           &                & WHIM & ($12\pm 5$)\%\\
           &                & Point Sources & $\geq$($35\pm 11$)\%\\
\enddata  
\end{deluxetable}

\end{document}